\documentclass[manuscript, nonacm]{acmart}

\renewcommand\footnotetextcopyrightpermission[1]{} 
\setcopyright{none} 
\acmConference{}{}{}
\acmBooktitle{}

\usepackage{graphicx}
\usepackage{amsmath,amsfonts}
\usepackage{textcomp}
\usepackage{color,soul}
\usepackage{multirow}
\usepackage{threeparttable}
\usepackage{csvsimple}
\usepackage{longtable}
\usepackage{caption}
\usepackage{bchart}
\usepackage[most]{tcolorbox}
\usepackage{makecell} 
\usepackage{array} 
\usepackage{listings}
\usepackage{xcolor}
\usepackage{enumitem}

\tcbset{
  mygraybox/.style={
    colback=gray!5,      
    colframe=black!15,   
    boxrule=0.4pt,       
    arc=1mm,             
    left=1mm, right=1mm, top=1mm, bottom=1mm, 
    fonttitle=\bfseries,
    coltitle=black,
  }
}

\newcommand{\secref}[1]{Section~\ref{sec:#1}}
\newcommand{\tabref}[1]{Table~\ref{tab:#1}}

\newcommand{\figref}[1]{Figure~\ref{fig:#1}}

\AtBeginDocument{%
  }

\setcopyright{acmlicensed}
\copyrightyear{2026}
\acmYear{2026}
\acmDOI{XXXXXXX.XXXXXXXX}
\acmConference[FORGE '2026]{ACM conference on AI Foundation Models and Software Engineering}{Rio de Janeiro, Brazil}

\acmISBN{978-1-4503-XXXX-X/2026/04}

\begin{document}

\title{Integrating Code Metrics into Automated Documentation Generation for Computational Notebooks}

\author{Mojtaba Mostafavi Ghahfarokhi}
\email{m.mostafavi@sharif.edu}
\affiliation{%
  \institution{Sharif University of Technology}
  \city{Tehran}
  \country{Iran}
}

\author{Hamed Jahantigh}
\email{hamed.jahantigh96@sharif.edu}
\affiliation{%
  \institution{Sharif University of Technology}
  \city{Tehran}
  \country{Iran}
}

\author{Alireza Asadi}
\email{alireza.asadi@sharif.edu}
\affiliation{%
  \institution{Sharif University of Technology}
  \city{Tehran}
  \country{Iran}
}

\author{Abbas Heydarnoori}
\email{heydarnoori@sharif.edu}
\affiliation{%
  \institution{Sharif University of Technology}
  \city{Tehran}
  \country{Iran}
}
\authornote{Abbas Heydarnoori is currently affiliated with the Department of Computer Science at Bowling Green State University, USA.}

\renewcommand{\shortauthors}{Mostafavi et al.}

\begin{abstract}
Effective code documentation is essential for collaboration, comprehension, and long-term software maintainability, yet developers often neglect it due to its repetitive nature. Automated documentation generation has evolved from heuristic and rule-based methods to neural network–based and large language model (LLM)–based approaches. However, existing methods often overlook structural and quantitative characteristics of code that influence readability and comprehension. Prior research suggests that code metrics capture information relevant to program understanding. Building on these insights, this paper investigates the role of source code metrics as auxiliary signals for automated documentation generation, focusing on computational notebooks, a popular medium among data scientists that integrates code, narrative, and results but suffers from inconsistent documentation. We propose a two-stage approach. First, the \emph{CodeSearchNet} dataset construction process was refined to create a specialized dataset from over 17 million code and markdown cells. After structural and semantic filtering, approximately 36,734 high-quality (code, markdown) pairs were extracted. Second, two modeling paradigms, a lightweight CNN–RNN architecture and a few-shot GPT-3.5 architecture, were evaluated with and without metric information. Results show that incorporating code metrics improves the accuracy and contextual relevance of generated documentation, yielding gains of 6\% in BLEU-1 and 3\% in ROUGE-L F1 for CNN–RNN-based architecture, and 9\% in BERTScore F1 for LLM-based architecture. These findings demonstrate that integrating code metrics provides valuable structural context, enhancing automated documentation generation across diverse model families.

\end{abstract}


\maketitle

\section{Introduction}
\label{sec:intro}
In modern software development, code documentation is essential for collaboration, maintenance, and knowledge transfer~\cite{xia2017measuring}. As projects grow in scale and teams become more distributed, clear documentation supports efficient comprehension and reduces onboarding effort~\cite{ghahfarokhi2025predicting}. However, developers often neglect documentation due to its repetitive nature and the effort required to keep it consistent with evolving codebases~\cite{oliveira2020evaluating}.

To reduce this burden, automated code documentation generation has been widely explored. Early methods based on static analysis and heuristics~\cite{sridhara2010towards, badihi2017crowdsummarizer} often ignored semantic and contextual cues. McBurney et al.~\cite{mcburney2014automatic} improved upon this by integrating Natural Language Generation (NLG)~\cite{reiter1997building} with SWUM~\cite{hill2009automatically} and PageRank-based importance measures~\cite{langville2006google}. Later, Neural Machine Translation (NMT) models~\cite{bahdanau2014neural} such as LSTM-based Seq2Seq architectures~\cite{hu2018deep, liu2018table} enhanced context awareness but required large datasets and high computational cost. More recent work employs large language models (LLMs) through prompt engineering~\cite{khan2022automatic, ahmed2022few, khan23automatic}, which improves flexibility but remains sensitive to prompt design and retrieval accuracy~\cite{shi2022evaluation, lu2024improving}.

Recent studies have also focused on integrating structural and semantic information to improve documentation quality. Approaches include Code2Seq models using Abstract Syntax Trees (ASTs)~\cite{alon2018code2seq}, API knowledge transfer for Java projects~\cite{hu2018summarizing}, and GraphCodeBERT’s use of Data Flow Graphs (DFGs)~\cite{guo2020graphcodebert}. LLM-based methods such as~\cite{ahmed2024automatic} further embed repository names, function signatures, and structural tags into prompts to enhance contextual relevance. Despite these advances, few approaches leverage code quality metrics as augmentation signals, even though such metrics strongly correlate with readability and comprehension~\cite{kasto2013measuring, posnett2011simpler, mostafavi2024distil}.

Software metrics play a central role in software measurement, enabling objective assessment of product quality throughout the development process~\cite{mathias1999role}. Source code metrics, such as Lines of Code (LOC), Lines of Comments (LOCom), Number of Operands (OPRND), or complexity measures, provide quantitative insights into structural properties and maintainability of code~\cite{nunez2017source}. Prior studies indicate that these metrics encapsulate information relevant to program comprehension~\cite{ghahfarokhi2025predicting, pimentel2019large, mostafavi2024can}. Motivated by this, the present study examines the impact of incorporating source code metrics into automated documentation generation models.

This analysis focuses on computational notebooks, which have become a preferred medium in data science due to their adherence to the literate programming paradigm~\cite{knuth1984literate,rule2018exploration,wenskovitch2019albireo}. Notebooks seamlessly integrate code, narrative, and results, yet documentation practices within them remain inconsistent~\cite{pimentel2019large,grotov2022large}. Many data scientists either lack formal software engineering training~\cite{wang2022documentation,muller2021data} or regard documentation as an unnecessary overhead~\cite{wang2020better}. Furthermore, the modular cell structure of notebooks naturally exposes a variety of code metrics~\cite{titov2022resplit,jiang2022elevating}, making them a suitable domain for evaluating metric-informed documentation generation methods. Accordingly, this study investigates whether information derived from code metrics can enhance automated documentation generation in computational notebooks. 

To address this problem, we propose a two-stage approach. In the first stage, a dedicated dataset was developed by refining and extending the CodeSearchNet data construction process~\cite{husain2019codesearchnet}. This effort began with the examination of over 17 million code and documentation cells, followed by rigorous structural and semantic analyses, which resulted in approximately 36,734 high-quality (code, markdown) pairs. In the second stage, two modeling paradigms were examined. A lightweight CNN–RNN architecture was first employed to observe the influence of metric information under constrained modeling capacity. Subsequently, a modern LLM, i.e., GPT‑3.5, which had limited prior exposure to the dataset used in this work, was evaluated in few-shot learning mode both with and without metric augmentation. The models generated documentation using the newly curated dataset through training and prompt engineering techniques. 
Experimental results indicate that augmenting input representations with code metrics substantially improves the accuracy and contextual relevance of the generated documentation, yielding gains of 6\% in BLEU‑1 and 3\% in ROUGE‑L F1 for neural models, and 9\% in BERTScore F1 for LLM-based models. These findings demonstrate that incorporating code metrics provides valuable structural context, effectively enhancing automated documentation generation across diverse model families. In summary, the main contributions of this paper are as follows:

\begin{itemize}[leftmargin=14pt]
    \item We introduce a new dataset for code documentation generation in computational notebooks, comprising approximately 36,734 (code, markdown) pairs.
    \item We investigate the impact of incorporating code metrics as an augmentation signal in deep learning models for documentation generation.
    \item We examine the influence of code metrics on documentation quality in LLM-based models through metric-informed prompting and evaluation.
\end{itemize}

The remainder of this paper is organized as follows. \secref{motivation} presents a motivating example that illustrates the impact of incorporating some code metrics into documentation generation. Next, \secref{codemetrics} describes the code metrics employed in this study. \secref{approach} details the dataset construction process and model architectures. \secref{eval} reports the evaluation results, while \secref{threads} summarizes the identified threats to validity. \secref{related_work} reviews related studies, and finally, \secref{conclusion} concludes the paper and outlines directions for future research.

\section{Motivating Example} \label{sec:motivation}
When LLMs are tasked with generating documentation for short and self-explanatory code cells, they often produce overly detailed, line-by-line explanations that differ markedly from how developers typically describe familiar operations. To illustrate this phenomenon, we examined an example in our dataset which demonstrates how a model may fail to capture the intended abstraction level.

\begin{lstlisting}[language=Python]
import pandas as pd
df = pd.read_csv("data.csv")
mean_value = df["age"].mean()
print(mean_value)
\end{lstlisting}

The model-generated documentation was:

\begin{tcolorbox}[mygraybox]
\textit{This code imports the pandas library, loads a CSV file named ``data.csv'', calculates the mean of the column 'age', and prints the result to the console.}
\end{tcolorbox}

Although technically correct, this description mirrors the code’s procedural flow rather than its purpose. In contrast, the corresponding human-authored markdown in the dataset was concise and goal-oriented:

\begin{tcolorbox}[mygraybox]
\textit{Calculate and display the average age from the dataset.}
\end{tcolorbox}

This disparity exemplifies a broader tendency of generative models to overlook structural cues that might signal code familiarity or simplicity. If a model could recognize that this cell is short, employs common APIs, and exhibits low cyclomatic complexity, it might infer that a succinct, purpose-level summary is more appropriate. To test this assumption, we selected another cell from the dataset that closely matches the first across multiple code metrics, including lines of code, average line length, number of parameters, comment word count, API familiarity, and overall complexity. The retrieved cell is shown below:

\begin{lstlisting}[language=Python]
import numpy as np
import matplotlib.pyplot as plt

values = np.random.normal(50, 10, 100)
plt.hist(values, bins=10)
plt.show()
\end{lstlisting}

Its associated human-written markdown was:

\begin{tcolorbox}[mygraybox]
\textit{Plot a histogram of randomly generated values.}
\end{tcolorbox}

This second cell was retrieved by our metric-based similarity module as the most relevant exemplar and then included in a one-shot prompt. When the model was prompted again to document the first code cell using this exemplar, the generated output became:

\begin{tcolorbox}[mygraybox]
\textit{Calculate and print the average age from the dataset.}
\end{tcolorbox}

This example demonstrates that metric-based similarity captures deeper structural relationships between code snippets beyond surface lexical overlap. By identifying exemplars with comparable complexity, API usage, and abstraction level, code metrics offer an interpretable and computationally efficient foundation for exemplar selection. This observation motivates our central hypothesis: incorporating explicit code metrics into exemplar retrieval and model conditioning can systematically improve the quality and relevance of LLM-generated notebook documentation.

\section{Code Metrics}
\label{sec:codemetrics}

\begin{table}[t]
\caption{Notebook Code Metrics}
\label{tab:code_metrics}
\centering
\begin{threeparttable}
\centering
\begin{tabular}{|c|c|}
\hline
\textbf{Metric} & \textbf{Abbreviation} \\
\hline\hline
\multicolumn{2}{|c|}{\textit{\textbf{i. Script-Based Metrics}}} \\ \hline\hline
\multicolumn{2}{|c|}{\textit{{i-a. Basic Metrics}}} \\ \hline
Number of Lines of Code & LOC \\ \hline
Number of Blank Lines of Code & BLC \\ \hline
Number of Lines of Comments & LOCom \\ \hline
Number of Comment Words & CW \\ \hline
Number of Statements & S \\ \hline
Number of Parameters & P \\ \hline
Number of User-Defined Functions & UDF \\ \hline
\multicolumn{2}{|c|}{\textit{{i-b. Complexity Metrics}}} \\ \hline
Nested Block Depth & NBD \\ \hline
Cyclomatic Complexity & CyC \\ \hline
Kind of Line of Code Identifier Density & KLCID \\ \hline
\multicolumn{2}{|c|}{\textit{{i-c. Halstead Metrics}}} \\ \hline
Number of Operands & OPRND \\ \hline
Number of Operators & OPRATOR \\ \hline
Number of Unique Operands & UOPRND \\ \hline
Number of Unique Operators & UOPRAT \\ \hline
\multicolumn{2}{|c|}{\textit{{i-d. Readability Metrics}}} \\ \hline
Average Line Length of Code & ALLC \\ \hline
Number of Identifiers & ID \\ \hline
Average Length of Identifiers & ALID \\ \hline\hline
\multicolumn{2}{|c|}{\textbf{\textit{ii. Notebook-Based Metrics}}} \\ \hline\hline
Number of Imports & I \\ \hline
External API Popularity & EAP \\ \hline
Number of Visualization Data Types & NDD \\ \hline
Executed Output & EC \\ \hline
\end{tabular}
\end{threeparttable}
\end{table}

Structural dimensions of code capture syntactic and organizational properties that are not fully reflected in semantic abstractions such as Abstract Syntax Trees (ASTs) or Data Flow Graphs (DFGs). To systematically investigate the influence of these structural indicators on models that generate markdown documentation from code, a suite of measurable \emph{code metrics} was compiled. 

Following the taxonomy introduced in~\cite{mostafavi2024distil}, 34 candidate metrics were identified from widely accepted code-analysis tools and studies. They were grouped into two primary categories:  
(i) \emph{Notebook-Based Metrics}, specifically designed for Jupyter notebook environments, and  
(ii) \emph{Script-Based Metrics}, describing general structural and lexical properties of Python code cells.  

Script-based metrics are divided into four subgroups:  
(i-a)~\emph{Basic Metrics}, capturing fundamental code quantities such as the numbers of statements, comments, and parameters;  
(i-b)~\emph{Complexity Metrics}, representing structural intricacy through measures like nested block depth or cyclomatic complexity;  
(i-c)~\emph{Halstead Metrics}, evaluating algorithmic complexity based on the frequency and diversity of operators and operands; and  
(i-d)~\emph{Readability Metrics}, reflecting code legibility in terms of line and identifier characteristics.

Because this research focuses on modeling code structure and semantics rather than markdown-only behavior, metrics pertaining exclusively to non-code markdown cells were excluded as irrelevant. After this refinement, 21 metrics were retained, comprising 17 Script-Based and 4 Notebook-Based features. The complete list of selected metrics is provided in \tabref{code_metrics}. These 21 features constitute the structural complement to the semantic similarity and human- (or LLM-) based evaluation criteria used in the benchmark study by~\cite{sun2025source}. As shown in that work, conventional automated metrics such as BLEU, METEOR, or ROUGE-L often fail to capture subtle stylistic and structural nuances in code summaries. By contrast, the inclusion of explicit structural indicators like those in \tabref{code_metrics} enables an evaluation pipeline to bridge this gap and reveal underlying code organization that purely text-based models may overlook. Notably, several of these metrics are intrinsically correlated with programming expertise and comprehension cost, offering high discriminative power even for state-of-the-art code comprehension models. Examples of particularly informative metrics include:

\begin{itemize}[leftmargin=11pt]
    \item \emph{External API Popularity (EAP):}  
    Quantifies how often the external libraries and APIs used in a notebook occur across the dataset. Following Scalabrino et al.~\cite{scalabrino2019automatically}, each API is assigned a popularity score proportional to its global frequency. The cumulative popularity of imports within a notebook yields its EAP score, which has been empirically linked to developer proficiency.

    \item \emph{Cyclomatic Complexity (CyC):}  
    Measures the number of linearly independent paths in the program's control-flow representation, as originally defined by Mathias~\cite{mathias1999role}. CyC provides an interpretable index of design intricacy and maintenance effort.
    
    \item \emph{Kind of Line of Code Identifier Density (KLCID):}  
    Reflects the density of unique identifiers across conceptually distinct lines of code, indicating cognitive load during code comprehension~\cite{klemola2003cognitive}. Higher values signify richer syntactic diversity and potentially greater interpretation cost.
\end{itemize}

\section{The Proposed Approach}
\label{sec:approach}
This section outlines the overall methodology developed to investigate the role of structural code metrics in improving automated documentation generation for computational notebooks. The approach is structured into three main parts: dataset construction, a baseline deep learning-based architecture, and an LLM-based architecture.

\subsection{Dataset Construction}
\label{sec:dataset}

The dataset construction process began with the DistilKaggle corpus~\cite{mostafavi2024distil}, which contains Kaggle Jupyter notebook cells, comprising more than 12 million code cells and over 5 million markdown cells between October 2020 and October 2023. The methodology for refining this corpus was organized into two main stages: \emph{structural analysis} and \emph{conceptual analysis}.

\begin{figure}[t]
  \centering
  \includegraphics[width=0.7\linewidth]{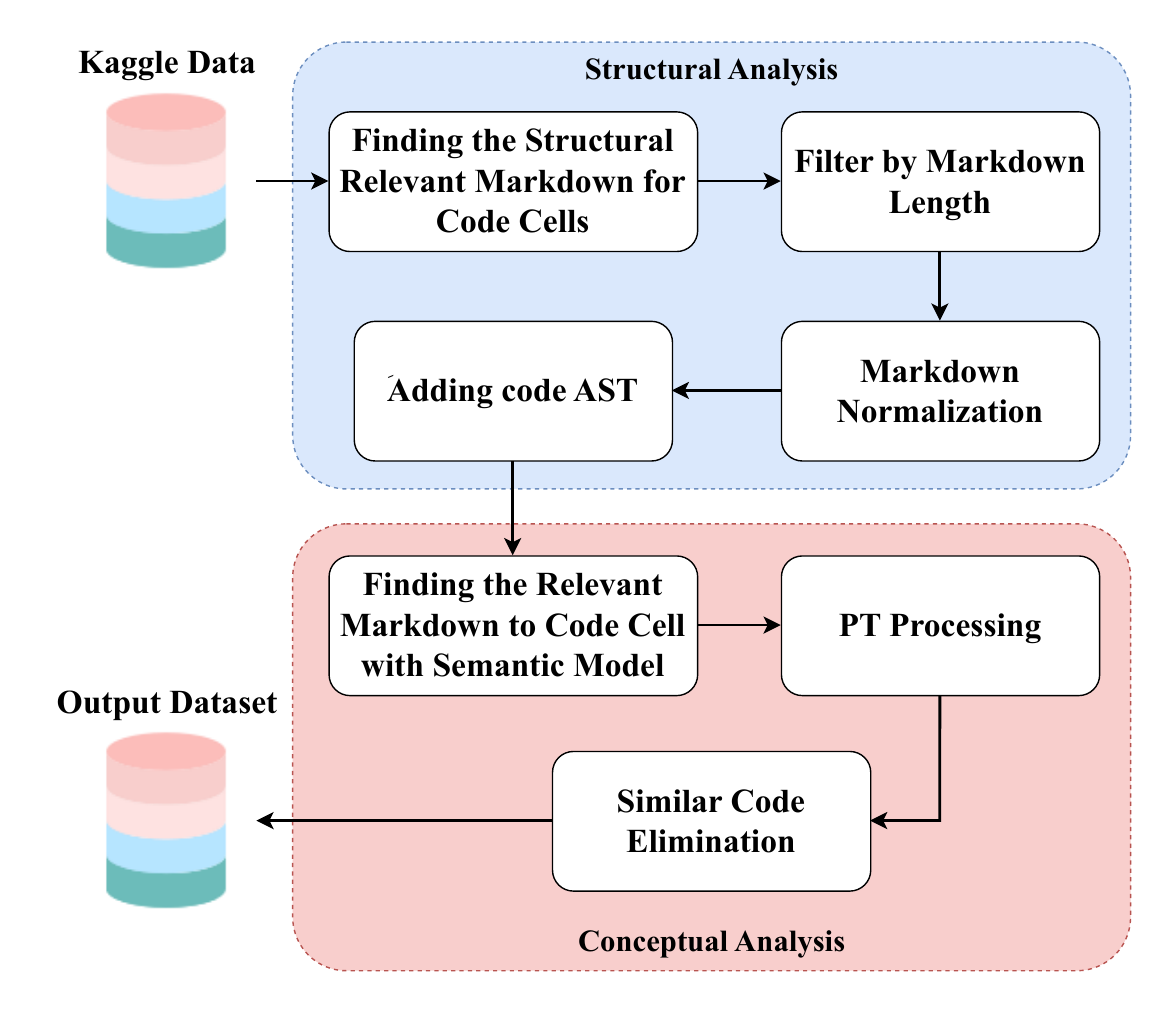}
  \caption{The Dataset Construction Methodology}
  \label{fig:methodology}
\end{figure}

In the structural analysis phase, individual code and markdown cells were examined to ensure compliance with specific structural criteria. This included filtering cells by function definitions, ensuring consistent (code, markdown) pairing, and imposing word limits on markdown text to avoid extreme length variations. The conceptual analysis phase applied broader quality checks, addressing factors such as user expertise, removal of duplicated content from forked notebooks, and verification of semantic alignment between code and markdown documentation. The overall process is illustrated in \figref{methodology}, with each phase detailed below.

\subsubsection{Structural Analysis} This phase includes the following sub-steps:

\begin{itemize}[leftmargin=11pt]
\item{\emph{Associating Structural Markdown:}}
A strict pairing criterion was applied to link each cell to its corresponding markdown. Only cells positioned between two markdown cells were considered, assuming that the preceding markdown serves as documentation for that single code cell. This avoids associating markdown with multiple subsequent cells. An additional semantic filtering step in \secref{conceptualFiltering} further verifies these associations. Applying the criterion resulted in 148,856 (code, markdown) pairs. To validate the pairing quality, a manual inspection of 100 randomly selected pairs was conducted, revealing that in 97\% of cases, the markdown accurately described the associated code cell.

\item{\emph{Markdown Length Filtering:}}
A markdown exceeding a certain token threshold can hinder the learning efficiency. An analysis of 3.3 million markdown cells showed an average length of 35 words (standard deviation 61). A maximum limit of 281 words, covering 99\% of the data, was set. The markdowns shorter than four words were removed as too minimal for meaningful learning. This filtering reduced the set to 118,340 (code, markdown) pairs.

\item{\emph{Markdown Normalization:}}
To improve training quality, markdown cells were cleaned by removing embedded code snippets, images, mathematical expressions, and HTML code, while preserving the original text in a separate column in the dataset for reference.
\end{itemize}

\subsubsection{Conceptual Analysis}
\label{sec:conceptualFiltering}
\begin{figure}[t]
  \centering
  \includegraphics[width=0.7\linewidth]{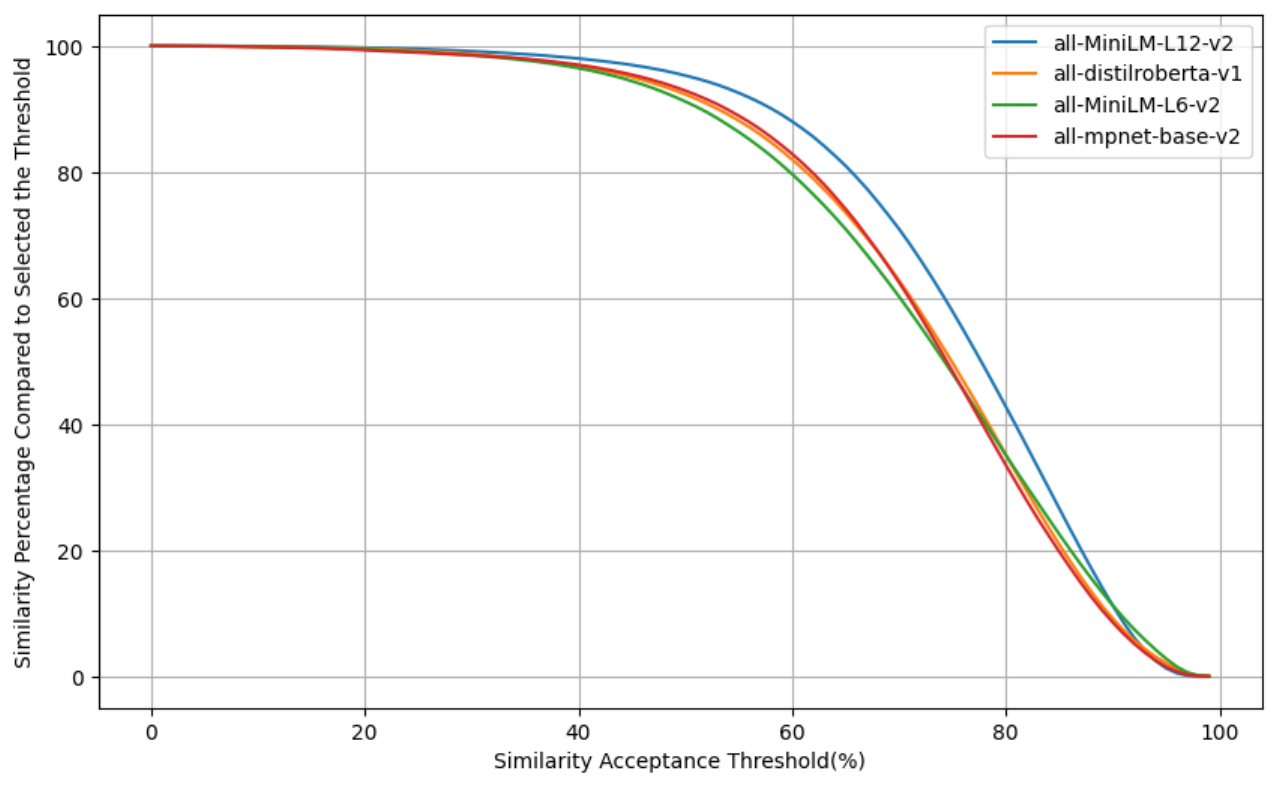}
  \caption{Model Comparison for Code and Documentation Similarity in the Ground Truth Dataset}
  \label{fig:embeddingmodel}
\end{figure}

This phase includes the following sub-steps:

\begin{itemize}[leftmargin=11pt]
\item{\emph{Semantic Relevance Filtering:}}
Despite passing structural checks, a markdown may not fully explain the associated code. To address this, sentence embedding models were used to measure the semantic similarity between each cell and its markdown. Four top-performing Sentence Transformer models, i.e., all-mpnet-base-v2, all-MiniLM-L6-v2, all-MiniLM-L12-v2, and all-distilroberta-v1, were evaluated using the Python subset of the CodeSearchNet dataset~\cite{husain2019codesearchnet}, which provides aligned (code, markdown) pairs. Cosine similarity scores were computed for each model, with \texttt{all-MiniLM-L12-v2} producing the highest scores (\figref{embeddingmodel}) and being selected for our dataset processing. A similarity threshold of 0.58 (corresponding to 90\% alignment confidence in the benchmark dataset) was adopted. Applying this filter retained 557,184 pairs from the original 2,587,963 candidates.

\item{\emph{Performance Tier Filtering:}}
As Kaggle assigns each user a Performance Tier (PT) score~\cite{alomar2021behind,scalabrino2019automatically}, we excluded notebooks authored by users with $PT=0$, which indicates limited experience. This filtering step removed approximately 33.4\% of the data, resulting in a final set of 371,071 pairs.

\item{\emph{Duplicate Code Removal:}}
Kaggle notebooks can be forked, often resulting in duplicated code~\cite{hernandez22scaling}. To preserve dataset diversity, only original, non-forked versions were retained. Facebook AI Similarity Search (Faiss)~\cite{johnson2019billion} was used to identify and exclude cells with more than 70\% code similarity.
\end{itemize}

Applying all filters to the initial corpus of over 17 million (code, markdown) pairs from 542,051 Kaggle notebooks (September 2015–October 2023) resulted in a final set of 36,734 high-quality (code, markdown) pairs. To assess quality, 100 random pairs (~5\% of the dataset) were reviewed by two experts, each with at least five years of Python experience. Pairs were labeled as ``Exact match,'' ``Strong match,'' ``Weak match,'' or ``Totally irrelevant'' following the criteria of Husain et~al.~\cite{husain2019codesearchnet}. The average distribution was 31\%, 52.5\%, 13\%, and 3.5\%, respectively. Cohen's kappa~\cite{cohen1960coefficient} was 0.85, indicating almost perfect inter-annotator agreement.

\subsection{Deep Learning–Based Architecture}
In the following, we first present the research question addressed in this phase and then describe the approach used to answer it.

\paragraph{\textbf{RQ1: Can code metrics enhance the performance of deep learning models in code documentation generation for computational notebooks?}}

By answering this question, we aim to determine whether the structural information provided by code metrics is useful for the documentation generation task. Our hypothesis is that simple deep learning models generally do not implicitly consider the structural information of code metrics. To test this, we conducted an experiment comparing the model's performance in two states: with plain code and with code augmented by metrics. By explicitly incorporating code metrics into the model, we aim to assess their impact on the performance of code documentation generation.

We chose the CNN-RNN model architecture as our candidate for a simple deep learning model. Kalchbrenner et al.~\cite{kalchbrenner2013recurrent} proposed this architecture, and their evaluations have demonstrated that using Convolutional Neural Networks (CNN) as encoders does not outperform Recurrent Neural Networks (RNN). Therefore, CNN-RNN is considered a simple deep learning model architecture for NLP tasks.

\begin{figure}[t]
  \centering
  \includegraphics[width=\linewidth]{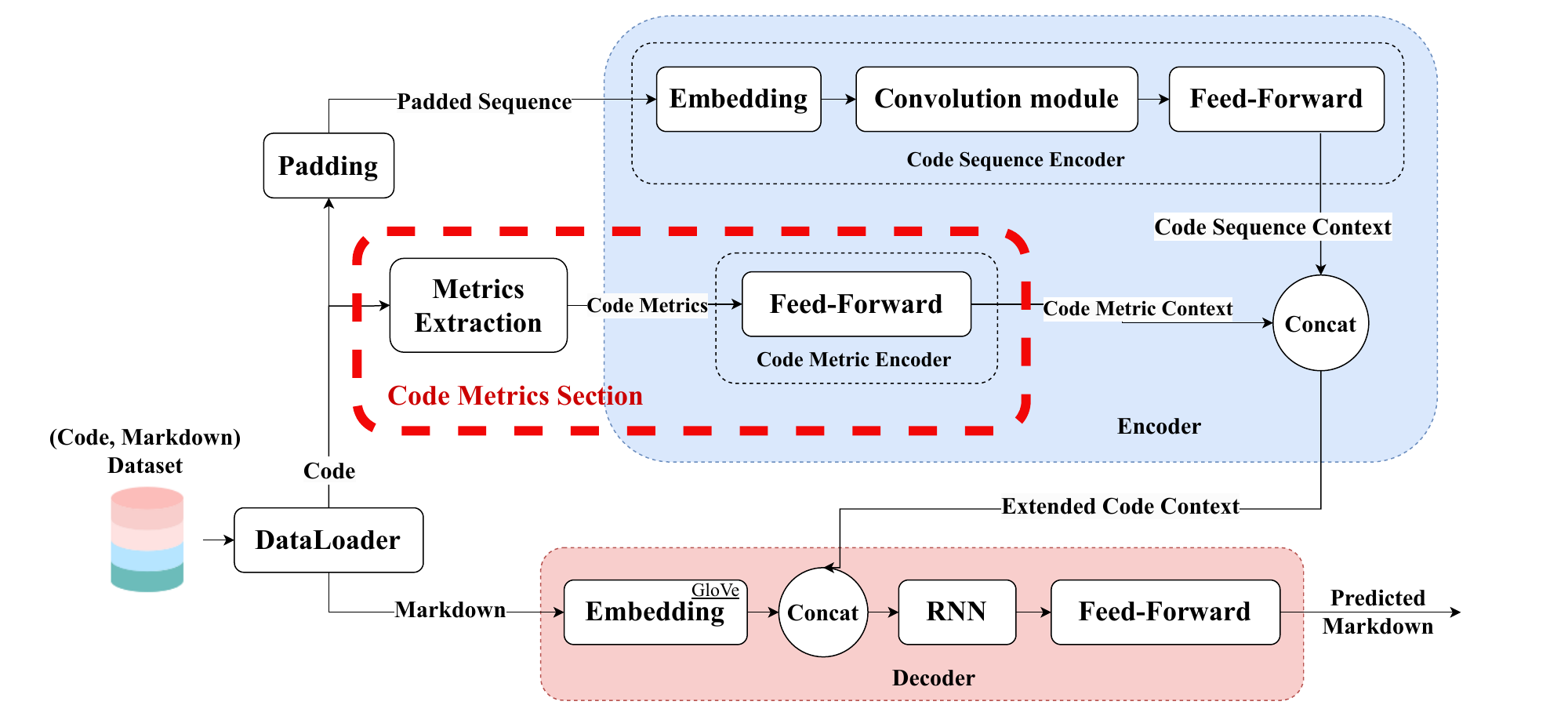}
  \caption{The CNN-RNN Architecture for Documentation Generation}
  \Description{It is a proposed model}
  \label{fig:cnnrnn-proposed}
\end{figure}

\figref{cnnrnn-proposed} illustrates the overall structure of this model, which consists of two primary components: the \textit{Encoder} and the \textit{Decoder}. The \textit{Encoder} begins with the Code Sequence Encoder segment, featuring custom word embeddings, a convolutional module, and a feed-forward network. The convolutional module comprises three layers of two-dimensional convolutional kernels paired with a max-pooling window. Following this, the feed-forward network includes a two-layer neural network, with each layer activated by a ReLU function. The context generated by the Encoder is integrated into the Decoder using the Par-inject method~\cite{tanti2018put}. The Decoder utilizes GloVe-based pre-trained word embeddings~\cite{pennington2014glove} and a lightweight RNN module. To enhance computational efficiency, the embedding vocabulary is pruned to include only terms found in the union of all markdowns in the training dataset. In addition, the feed-forward module within the Decoder is a single-layer neural network with a ReLU activation function. It is important to note that the highlighted \textit{Code Metrics} section is not included in this simple CNN-RNN architecture. Instead, it is integrated into the model which is explained below.

The \textit{Code Metrics} component is divided into two phases: \textit{Metrics Extraction} and the \textit{Code Metrics Encoder}. In the \textit{Metrics Extraction} phase, we gather a variety of metrics inspired by previous research, including cell complexity and function usage~\cite{grotov2022large}, API popularity~\cite{8651396}, the distribution of markdown cells or import statements~\cite{pimentel2019large}, and Halstead Metrics~\cite{posnett2011simpler}. A code feature matrix is generated from the aggregated metrics and fed into the \textit{Code Metrics Encoder}. This feature matrix is further contextualized using a simple feed-forward module. Finally, the code metrics context is combined with the code sequence context from the \textit{Code Sequence Encoder}, forming the comprehensive encoder context. \secref{evaluation-cnnrnn} details the evaluation results for this architecture.

\subsection{LLM-Based Architecture}
\label{llm_approach}
To investigate the impact of code metrics on LLM-based architectures, we first formulate two research questions and then describe our approach in detail.

\paragraph{\textbf{RQ2: How can code metrics be effectively integrated into LLM prompts to enhance the accuracy, relevance, and fluency of generated code documentation?}}

Building upon insights from RQ1, this question investigates the design space of augmentation strategies for LLMs, focusing on methods to embed or align code metric information within prompts or input representations. To explore this, we conducted experiments using GPT-3.5\footnote{\url{https://platform.openai.com/docs/models/gpt-3.5}} as the primary evaluation model, chosen for its modern architecture and minimal prior exposure to our dataset. Using paired (code, markdown) samples, we designed tailored prompts for GPT-3.5 under both baseline and metric-augmented conditions. The resulting documentation was evaluated using established metrics for accuracy, relevance, and fluency~\cite{shi2022evaluation}. Our analysis identifies the augmentation technique that delivers the greatest performance improvement while preserving generation stability.

\paragraph{\textbf{RQ3: Can code metrics enhance the performance of LLMs in documentation generation for computational notebooks?}}

This question examines the potential of metric augmentation in state-of-the-art transformer-based models for this domain. Rather than conducting additional experiments, the findings from RQ2, which include both baseline and augmented results, directly inform this question. The results help determine whether improvements seen in LLMs and whether metric-informed prompts could be a consistent method for boosting notebook documentation quality across different architectures~\cite{ahmed2022few,lu2024improving}.

\begin{figure}[t]
  \centering
  \includegraphics[width=\linewidth]{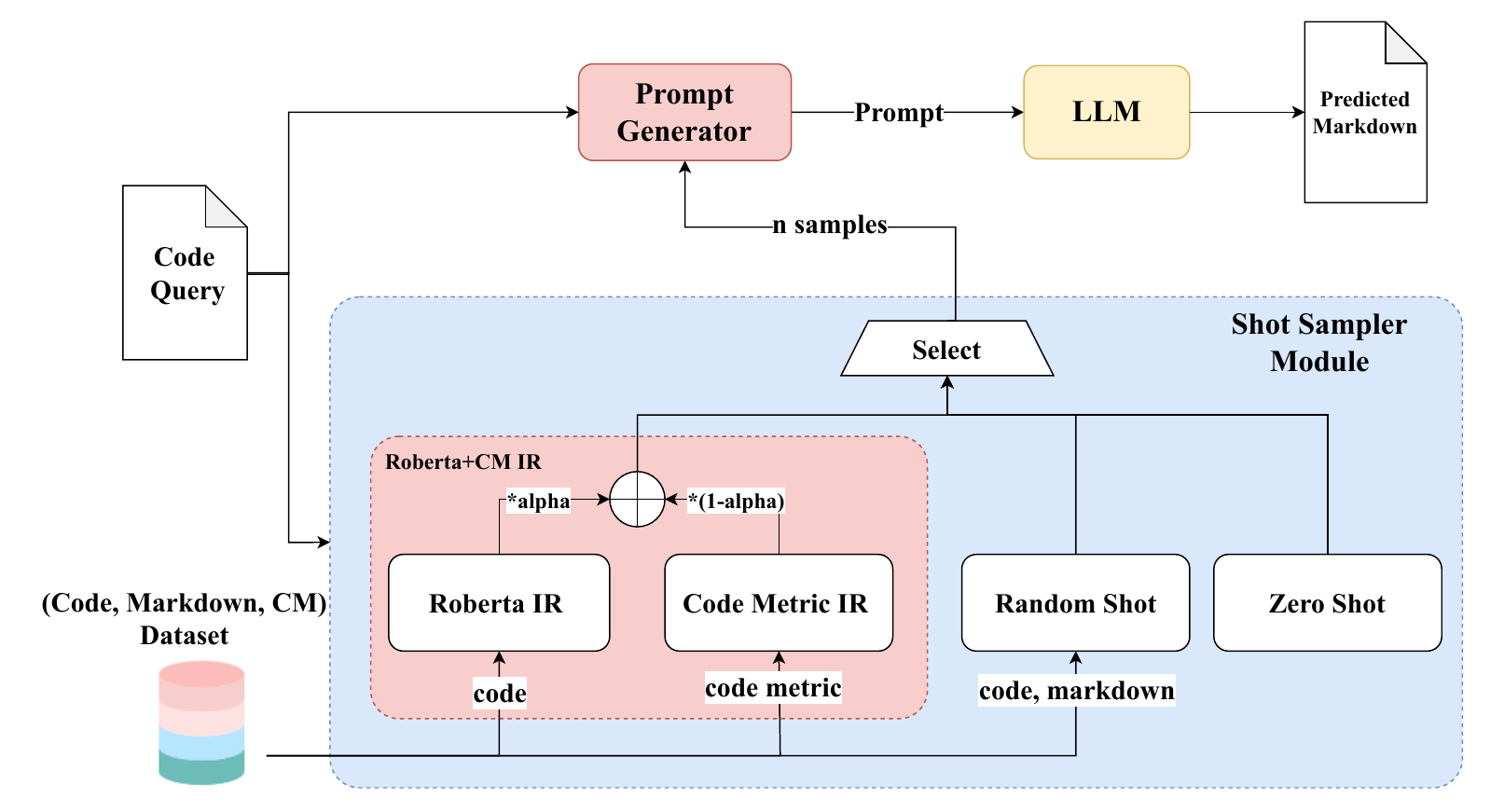}
  \caption{The Few-shot LLM-based Architecture for Documentation Generation}
  \label{fig:fewshot-architecture}
\end{figure}

Few-shot learning approaches for code-related tasks have gained significant popularity~\cite{nashid2023retrieval}, owing to their accuracy, generality, and the advantage of avoiding costly model training. Accordingly, we adopt this paradigm for code documentation generation. The overall architecture of our approach is illustrated in \figref{fewshot-architecture}. It consists of three key components: the \textit{Shot Sampler}, the \textit{Prompt Generator}, and the \textit{LLM} module. The \textit{Shot Sampler} selects example samples to be utilized by the \textit{Prompt Generator}. Notably, the design imposes no constraints on the number of samples, allowing the module to return even zero shots when appropriate.

As noted in prior few-shot learning studies~\cite{ahmed2022few,zhou2023towards,bravzinskas2020few}, the \textit{Shot Sampler} module commonly incorporates an IR component, such as BM25~\cite{robertson1996okapi} or RoBERTa-based retrieval models~\cite{nashid2023retrieval}, to identify the most similar examples and thereby maximize the informational relevance for the LLM. The number of retrieved samples (\textit{n}) is typically set to 0, 1, 5, or 10, corresponding to zero-shot, one-shot, and few-shot learning configurations, respectively. It is important to note that retrieval is performed solely based on code similarity, without considering the associated markdown. For our evaluations (see \secref{evaluation-llm}), we employ a diverse set of shot samplers:

\begin{itemize}[leftmargin=11pt]
    \item \emph{Zero-Shot:} This sampler always returns an empty list of samples.

    \item \emph{Random-Shot:} This sampler selects a random list of (code, markdown) pairs from the training dataset with a specified length for each input.

    \item \emph{Roberta~IR:} This sampler uses an IR module to find the most similar samples to a given code query. It employs RoBERTa~\cite{nashid2023retrieval} embeddings\footnote{\url{https://huggingface.co/flax-sentence-embeddings/st-codesearch-distilroberta-base}} to compute similarity scores between code pairs. For this sampler, $alpha$ in \figref{fewshot-architecture} is set to~1.

    \item \emph{Code Metric IR (CM~IR):} This unique sampler is designed specifically for code metrics. It uses an IR module to select (code, markdown) pairs from our dataset that have code metrics similar to the code query. The similarity is calculated using cosine similarity~\cite{singhal2001modern}, as the code metrics have a vector nature. For this sampler, $alpha$ shown in \figref{fewshot-architecture} is~0. This module is one of our candidate IR modules utilized in few-shot approaches.

    \item \emph{Roberta+CM~IR:} This sampler combines the Roberta and CM IR methods. It features an IR that calculates similarity scores between codes using both Roberta and CM~IR methods. The overall similarity score is obtained by averaging the two scores on a percentage scale($alpha=0.5$). Moreover, for general use, we can employ a weighted average with the hyper-parameter $alpha$. This approach aims to push the boundaries of state-of-the-art accuracy by optimizing the integration of the two scoring methods.
\end{itemize}

The \textit{Prompt Generator} module is responsible for generating prompts based on the code query and shot samples. Typically, prompt generation tasks are template-driven. Additionally, various augmentations can be applied to the generated prompt, such as incorporating AST, DFG, and code metrics~\cite{ahmed2024automatic}. The \textit{Prompt Generator} modules we use follow a specific template, but the content within the prompts falls into two categories:

\begin{itemize}[leftmargin=11pt]
    \item \emph{Without Metric Type:} This prompt contains the code query and the code and markdown of the samples.
    \item \emph{With Metric Type:} This prompt includes the code and metrics of the query, as well as the code, metrics, and markdown of the samples. Although this module uses the same input as the previous type, it includes a \textit{Metrics Extraction} module that calculates the metrics and incorporates them into the generated prompt.
\end{itemize}

\begin{table}[t]
    \caption{BLEU Evaluation Results for CNN-RNN Models: A Comparison Between the Base and Metric-Augmented (MA) Variants. P-values were computed using the Wilcoxon signed-rank test.}
    \setlength\tabcolsep{3pt}
    \begin{tabular}{|c|c|c|c|c|c|c|c|c|}
        \hline
         & \multicolumn{2}{|c|}{BLEU-1}&  \multicolumn{2}{|c|}{BLEU-2}&  \multicolumn{2}{|c|}{BLEU-3}& \multicolumn{2}{|c|}{BLEU-4} \\ \hline
        Fold & Base&MA & Base&MA & Base&MA & Base&MA \\ \hline
        1 & 0.27& 0.34 & 0.11& 0.18 & 0.05& 0.14 & 0.03& 0.12 \\ \hline
        2 & 0.27& 0.33 & 0.11& 0.18 & 0.05& 0.14 & 0.03& 0.12 \\ \hline
        3 & 0.25& 0.34 & 0.09& 0.19 & 0.04& 0.15 & 0.02& 0.13 \\ \hline
        4 & 0.23& 0.34 & 0.09& 0.19 & 0.04& 0.15 & 0.02& 0.13 \\ \hline
        5 & 0.29& 0.25 & 0.11& 0.11 & 0.05& 0.08 & 0.03& 0.06 \\ \hline
        6 & 0.28& 0.34 & 0.11& 0.2 & 0.05& 0.16 & 0.03& 0.14 \\ \hline
        7 & 0.28& 0.34 & 0.12& 0.2 & 0.06& 0.15 & 0.04& 0.14 \\ \hline
        8 & 0.24& 0.31 & 0.08& 0.17 & 0.05& 0.13 & 0.03& 0.11 \\ \hline
        9 & 0.27& 0.33 & 0.11& 0.19 & 0.05& 0.14 & 0.03& 0.13 \\ \hline
        10 & 0.27& 0.28 & 0.1& 0.11 & 0.05& 0.05 & 0.03& 0.03 \\ \hline
        \hline
        avg & 0.26& 0.32 & 0.1& 0.17 & 0.05& 0.13 & 0.03& 0.11 \\ \hline
        avg-gain & \multicolumn{2}{|c|}{+0.06} & \multicolumn{2}{|c|}{+0.07} & \multicolumn{2}{|c|}{+0.08} & \multicolumn{2}{|c|}{+0.08} \\ \hline
        std & 0.02& 0.03 & 0.01& 0.03 & 0.01& 0.04 & 0.005& 0.04 \\ \hline
        \hline
        p-value & \multicolumn{2}{|c|}{<0.01} & \multicolumn{2}{|c|}{<0.01} & \multicolumn{2}{|c|}{<0.01} & \multicolumn{2}{|c|}{<0.01} \\ \hline
    \end{tabular}
    \label{tab:cnn_rnn_evaluation_result_bleu}
\end{table}

\begin{table*}[t]
    \caption{ROUGE Evaluation Results for CNN-RNN Models: A Comparison Between the Base and Metric-Augmented (MA) Variants. P-values were computed using the Wilcoxon signed-rank test.}
    \setlength\tabcolsep{3pt}
    \begin{tabular}{|c|c|c|c|c|c|c|c|c|c|c|c|c|c|c|c|c|c|c|}
        \hline
         & \multicolumn{6}{|c|}{ROUGE-1}&  \multicolumn{6}{|c|}{ROUGE-2}&  \multicolumn{6}{|c|}{ROUGE-L} \\ \hline
        & \multicolumn{2}{|c|}{F1} & \multicolumn{2}{|c|}{Precision} & \multicolumn{2}{|c|}{Recall} & \multicolumn{2}{|c|}{F1} & \multicolumn{2}{|c|}{Precision} & \multicolumn{2}{|c|}{Recall} & \multicolumn{2}{|c|}{F1} & \multicolumn{2}{|c|}{Precision} & \multicolumn{2}{|c|}{Recall} \\ \hline
        Fold & Base&MA & Base&MA & Base&MA & Base&MA & Base&MA & Base&MA & Base&MA & Base&MA & Base&MA \\ \hline
        1 & 0.32 & 0.36 & 0.31 & 0.38 & 0.34 & 0.38 & 0.02 & 0.08 & 0.02 & 0.08 & 0.02 & 0.08 & 0.32 & 0.36 & 0.31 & 0.38 & 0.34 & 0.37 \\ \hline
        2 & 0.32 & 0.36 & 0.31 & 0.37 & 0.35 & 0.38 & 0.02 & 0.08 & 0.02 & 0.08 & 0.02 & 0.08 & 0.32 & 0.36 & 0.31 & 0.37 & 0.34 & 0.38 \\ \hline
        3 & 0.29 & 0.36 & 0.27 & 0.38 & 0.33 & 0.38 & 0.02 & 0.08 & 0.02 & 0.08 & 0.02 & 0.08 & 0.28 & 0.36 & 0.26 & 0.38 & 0.33 & 0.38 \\ \hline
        4 & 0.28 & 0.36 & 0.25 & 0.37 & 0.34 & 0.38 & 0.02 & 0.07 & 0.02 & 0.07 & 0.02 & 0.07 & 0.28 & 0.35 & 0.25 & 0.37 & 0.34 & 0.38 \\ \hline
        5 & 0.39 & 0.3 & 0.54 & 0.31 & 0.31 & 0.33 & 0.02 & 0.03 & 0.03 & 0.03 & 0.02 & 0.03 & 0.39 & 0.3 & 0.54 & 0.31 & 0.31 & 0.33 \\ \hline
        6 & 0.37 & 0.36 & 0.52 & 0.37 & 0.3 & 0.38 & 0.01 & 0.08 & 0.02 & 0.08 & 0.01 & 0.08 & 0.37 & 0.36 & 0.52 & 0.37 & 0.3 & 0.38 \\ \hline
        7 & 0.37 & 0.36 & 0.53 & 0.37 & 0.3 & 0.38 & 0.01 & 0.08 & 0.02 & 0.08 & 0.01 & 0.08 & 0.37 & 0.36 & 0.53 & 0.37 & 0.3 & 0.38 \\ \hline
        8 & 0.3 & 0.35 & 0.25 & 0.36 & 0.39 & 0.37 & 0.02 & 0.06 & 0.02 & 0.06 & 0.03 & 0.06 & 0.29 & 0.35 & 0.25 & 0.36 & 0.38 & 0.37 \\ \hline
        9 & 0.32 & 0.35 & 0.31 & 0.37 & 0.35 & 0.37 & 0.02 & 0.07 & 0.02 & 0.07 & 0.02 & 0.07 & 0.31 & 0.35 & 0.31 & 0.37 & 0.34 & 0.37 \\ \hline
        10 & 0.32 & 0.32 & 0.31 & 0.35 & 0.35 & 0.31 & 0.02 & 0.02 & 0.02 & 0.02 & 0.02 & 0.02 & 0.32 & 0.32 & 0.31 & 0.35 & 0.35 & 0.31 \\ \hline
        \hline
        avg & 0.33 & 0.35 & 0.36 & 0.36 & 0.34 & 0.36 & 0.02 & 0.06 & 0.02 & 0.07 & 0.02 & 0.06 & 0.32 & 0.35 & 0.36 & 0.36 & 0.33 & 0.36 \\ \hline
        avg-gain & \multicolumn{2}{|c|}{+0.02} & \multicolumn{2}{|c|}{$\approx 0$} & \multicolumn{2}{|c|}{+0.02} & \multicolumn{2}{|c|}{+0.04} & \multicolumn{2}{|c|}{+0.05} & \multicolumn{2}{|c|}{+0.04} & \multicolumn{2}{|c|}{+0.03} & \multicolumn{2}{|c|}{$\approx 0$} & \multicolumn{2}{|c|}{+0.03} \\ \hline
        std & 0.04 & 0.02 & 0.12 & 0.02 & 0.03 & 0.02 & 0 & 0.02 & 0 & 0.02 & 0.01 & 0.02 & 0.04 & 0.02 & 0.12 & 0.02 & 0.03 & 0.02 \\ \hline
        \hline
        p-value & \multicolumn{2}{|c|}{0.14} & \multicolumn{2}{|c|}{0.5} & \multicolumn{2}{|c|}{0.02} & \multicolumn{2}{|c|}{<0.01} & \multicolumn{2}{|c|}{<0.01} & \multicolumn{2}{|c|}{<0.01} & \multicolumn{2}{|c|}{0.12} & \multicolumn{2}{|c|}{0.5} & \multicolumn{2}{|c|}{0.02} \\ \hline
    \end{tabular}
    \label{tab:cnn_rnn_evaluation_result_rouge}
\end{table*}

\section{Evaluation}
\label{sec:eval}
In this section, we evaluate the proposed approaches and answer the research questions based on the experimental results.

\subsection{Experiment Design}

\subsubsection{Experiment Setup}
\label{sec:experiment-setup}
For our deep learning-based architecture, we evaluated two models: \textit{CNN-RNN} and \textit{CNN-RNN Metric-Augmented (MA)}. All experimental phases (training, testing, and evaluation) were conducted in a Kaggle environment with a free subscription tier, equipped with dual T4 GPUs (15~GB memory each), 73.1~GB of disk space, and 29~GB of RAM. This configuration provided a robust yet accessible platform for our computational requirements. Both models were implemented in PyTorch and trained for 200 epochs per experiment. The key hyperparameters were set as follows: learning rate = $3\times10^{-4}$, RNN hidden size = 512, and encoder context size = 246.

For our LLM-based architecture, we used GPT-3.5 as the primary model for all evaluations. All experiments were conducted in the same Kaggle environment described above. We accessed GPT-3.5 through the OpenAI API\footnote{\url{https://platform.openai.com/docs/models/gpt-3.5}} using the OpenAI Python library, with the temperature parameter set to 0.5 and default values for all other parameters\footnote{\url{https://platform.openai.com/docs/api-reference/chat}}. Prompt engineering followed OpenAI's recommended best practices\footnote{\url{https://help.openai.com/en/articles/6654000-best-practices-for-prompt-engineering-with-the-openai-api}}, incorporating structured instructions at the beginning and using triple quotes to separate instructions from the context. Few-shot learning prompts were employed to demonstrate the expected output format through representative examples.

The CNN-RNN and LLM-based models are included in our replication package~\cite{ReplicationPackage}. These models provide a diverse set of methods for generating and evaluating code documentation within our experimental framework.

\subsubsection{Evaluation Methods}

For our deep learning-based approach, we have performed a 10-fold cross validation on our gathered dataset, using the scikit-learn library\footnote{\url{https://scikit-learn.org/stable/}}, to reduce the effect of randomness on the results. For each fold, we have split the data by 8, 1, 1 for training, testing, and evaluation stages, respectively. Moreover, we have used the Wilcoxon signed-rank test with $\alpha=5\%$~\cite{wilcoxon1946individual} as a statistical test to validate the superiority of the proposed model against the baseline model with respect to randomness.

For the LLM-based approach, a dedicated test set containing 10\% of the 36,734 total (code, markdown) pairs was constructed to provide a diverse and representative basis for evaluation. The test set was generated using stratified sampling across key categories to preserve the overall data distribution. To maintain experimental consistency and comparability across models, all evaluations employed this same test set. The remaining (code, markdown) pairs were allocated for training, from which the \emph{Shot Sampler} module dynamically selected examples during the few-shot construction process.

\subsubsection{Metrics Used}

We employed BLEU~\cite{papineni2002bleu} and ROUGE~\cite{lin2004rouge} metrics for experiments involving the deep learning models. In addition, for the LLM experiments, we used BERTScore~\cite{zhang2019bertscore}, a semantic similarity–based evaluator that more effectively captures contextual and semantic relationships. These metrics are widely adopted in NLP research and provide standardized measures for evaluating the quality of machine-generated text.

BLEU~\cite{papineni2002bleu} measures lexical overlap between generated and reference texts, while ROUGE~\cite{lin2004rouge} evaluates n-gram and sequence-level similarity across reference sets. Together, these metrics provide a comprehensive view of model performance. In contrast, BERTScore~\cite{zhang2019bertscore} leverages contextual embeddings from BERT to assess semantic similarity between generated and reference texts. Considering the advanced conceptual generation capabilities of GPT-3.5, we additionally employed BERTScore to more effectively capture its semantic performance.

\subsection{Experiment Results}
In this section, we analyze and discuss the results obtained from our experiments.

\subsubsection{Experiment Results for Deep Learning–Based Architecture}\label{sec:evaluation-cnnrnn}
We present the BLEU scores for each fold of both the baseline and the proposed CNN-RNN models in \tabref{cnn_rnn_evaluation_result_bleu}. The average BLEU-1 score of the baseline model across all folds is 26\%, which is 6\% lower than that of the proposed model. Notably, the baseline model surpasses the proposed model in BLEU-1 for fold~\#5, while their scores are nearly identical in fold~\#10. In the remaining folds, the proposed model achieves substantially higher BLEU-1 scores. Furthermore, the proposed model consistently outperforms the baseline model in BLEU-2, BLEU-3, and BLEU-4 across most folds. On average, the proposed model achieves improvements of 7\%, 8\%, and 8\% in BLEU-2, BLEU-3, and BLEU-4, respectively. The baseline model's average scores are notably lower across these metrics. The superiority of the proposed model is further supported by the Wilcoxon signed-rank test, which yielded a p-value of $<0.01$ for all BLEU metrics. This result indicates that our model consistently generates text that is more similar to the reference outputs.

Similarly, we present the ROUGE scores for each fold of both the baseline and the proposed CNN-RNN models in \tabref{cnn_rnn_evaluation_result_rouge}. \tabref{cnn_rnn_evaluation_result_rouge} reports the ROUGE-1, ROUGE-2, and ROUGE-L scores, including F1, precision, and recall values for each fold, for both the baseline (Base) and the metric-augmented (MA) models. Overall, the metric-augmented model generally achieves higher scores across all ROUGE metrics compared to the baseline model. In particular, the MA model outperforms the baseline in terms of F1, precision, and recall, as evidenced by the higher average scores. Furthermore, the Wilcoxon signed-rank test results, shown in the table, indicate statistically significant differences between the two models across several ROUGE metrics, with p-values less than 0.05.

In conclusion, the results suggest that code metrics capture structural information that is beneficial for code documentation generation in computational notebooks, as evidenced by the superior BLEU and ROUGE scores.

\begin{tcolorbox}
\textbf{RQ1:} \textit{Can code metrics enhance the performance of deep learning models in code documentation generation for computational notebooks?}
\par \noindent
\textbf{Answer:} Code metrics positively influence documentation generation for deep learning models in computational notebooks. In our experiments, incorporating these metrics resulted in consistent average improvements across all BLEU and ROUGE scores.
\end{tcolorbox}


\begin{table}[t]
    \caption{BLEU Evaluation Results for LLM Experiments Conducted with the GPT‑3.5 Model}
    \begin{tabular}{|c|c|c|c|c|c|}
        \hline
        Prompt Generator & Shot Sampler & BLEU-1 & BLEU-2 & BLEU-3 & BLEU-4 \\ \hline
        \multirow{5}{*}{No Metric} & Zero‑Shot & 0.112 & 0.046 & 0.022 & 0.012 \\ \cline{2-6}
        ~ & Random‑Shot & 0.141 & 0.062 & 0.032 & 0.019 \\ \cline{2-6}
        ~ & Roberta~IR & 0.226 & 0.165 & 0.140 & 0.126 \\ \cline{2-6}
        ~ & CM~IR & 0.230 & 0.168 & 0.143 & 0.130 \\ \cline{2-6}
        ~ & Roberta+CM~IR & 0.233 & 0.171 & 0.146 & 0.132 \\ \hline
        \multirow{5}{*}{With Metric} & Zero‑Shot & 0.106 & 0.042 & 0.020 & 0.011 \\ \cline{2-6}
        ~ & Random‑Shot & 0.142 & 0.061 & 0.032 & 0.018 \\ \cline{2-6}
        ~ & Roberta~IR & 0.221 & 0.157 & 0.131 & 0.118 \\ \cline{2-6}
        ~ & CM~IR & 0.207 & 0.142 & 0.115 & 0.102 \\ \cline{2-6}
        ~ & Roberta+CM~IR & 0.225 & 0.161 & 0.135 & 0.121 \\ \hline
    \end{tabular}
    \label{tab:fewshot_evaluation_result_gpt3_bleu}
\end{table}

\begin{table*}[t]
    \caption{ROUGE Evaluation Results for LLM Experiments Conducted with the GPT‑3.5 Model}
    \setlength\tabcolsep{3pt}
    \begin{tabular}{|c|c|c|c|c|c|c|c|c|c|c|}
        \hline
        \multirow{2}{*}{Prompt Generator} & \multirow{2}{*}{Shot Sampler} & \multicolumn{3}{|c|}{ROUGE-1} & \multicolumn{3}{|c|}{ROUGE-2} & \multicolumn{3}{|c|}{ROUGE-L} \\ \cline{3-11}
        ~ & ~ & F1 & Precision & Recall & F1 & Precision & Recall & F1 & Precision & Recall \\ \hline
        \multirow{5}{*}{No Metric} & Zero‑Shot & 0.184 & 0.147 & 0.358 & 0.041 & 0.032 & 0.089 & 0.137 & 0.107 & 0.284 \\ \cline{2-11}
        ~ & Random‑Shot & 0.213 & 0.241 & 0.287 & 0.056 & 0.065 & 0.078 & 0.167 & 0.188 & 0.230 \\ \cline{2-11}
        ~ & Roberta~IR & 0.296 & 0.318 & 0.363 & 0.152 & 0.157 & 0.176 & 0.257 & 0.272 & 0.316 \\ \cline{2-11}
        ~ & CM~IR & 0.292 & 0.313 & 0.356 & 0.147 & 0.151 & 0.168 & 0.251 & 0.266 & 0.307 \\ \cline{2-11}
        ~ & Roberta+CM~IR & 0.297 & 0.318 & 0.358 & 0.151 & 0.156 & 0.173 & 0.257 & 0.274 & 0.311 \\ \hline
        \multirow{5}{*}{With Metric} & Zero‑Shot & 0.175 & 0.141 & 0.335 & 0.036 & 0.028 & 0.075 & 0.130 & 0.103 & 0.265 \\ \cline{2-11}
        ~ & Random‑Shot & 0.211 & 0.229 & 0.290 & 0.053 & 0.057 & 0.077 & 0.164 & 0.176 & 0.232 \\ \cline{2-11}
        ~ & Roberta~IR & 0.292 & 0.304 & 0.364 & 0.144 & 0.146 & 0.169 & 0.250 & 0.257 & 0.314 \\ \cline{2-11}
        ~ & CM~IR & 0.278 & 0.291 & 0.354 & 0.130 & 0.134 & 0.155 & 0.236 & 0.245 & 0.302 \\ \cline{2-11}
        ~ & Roberta+CM~IR & 0.295 & 0.309 & 0.367 & 0.147 & 0.150 & 0.172 & 0.254 & 0.264 & 0.316 \\ \hline
    \end{tabular}
    \label{tab:fewshot_evaluation_result_gpt3_rouge}
\end{table*}

\begin{table}[t]
    \caption{BERTScore Evaluation Results for LLM Experiments Conducted with the GPT‑3.5 Model}
    \setlength\tabcolsep{3pt}
    \begin{tabular}{|c|c|c|c|c|}
        \hline
        \multirow{2}{*}{Prompt Generator} & \multirow{2}{*}{Shot Sampler} & \multicolumn{3}{|c|}{BERTScore} \\ \cline{3-5}
        ~ & ~ & F1 & Precision & Recall \\ \hline
        \multirow{5}{*}{No Metric} & Zero‑Shot & 0.234 & 0.182 & 0.302 \\ \cline{2-5}
        ~ & Random‑Shot & 0.246 & 0.220 & 0.282 \\ \cline{2-5}
        ~ & Roberta~IR & 0.313 & 0.282 & 0.355 \\ \cline{2-5}
        ~ & CM~IR & 0.309 & 0.279 & 0.349 \\ \cline{2-5}
        ~ & Roberta+CM~IR & 0.312 & 0.283 & 0.352 \\ \hline
        \multirow{5}{*}{With Metric} & Zero‑Shot & 0.217 & 0.167 & 0.283 \\ \cline{2-5}
        ~ & Random‑Shot & 0.248 & 0.223 & 0.282 \\ \cline{2-5}
        ~ & Roberta~IR & 0.317 & 0.288 & 0.356 \\ \cline{2-5}
        ~ & CM~IR & 0.304 & 0.277 & 0.342 \\ \cline{2-5}
        ~ & Roberta+CM~IR & 0.320 & 0.293 & 0.357 \\ \hline
    \end{tabular}
    \label{tab:fewshot_evaluation_result_gpt3_bert}
\end{table}

\subsubsection{Experiment Results for LLM-Based Architecture}\label{sec:evaluation-llm}

\begin{figure*}[t]
  \centering
  \begin{minipage}[b]{0.32\linewidth}
    \centering
    \includegraphics[width=\linewidth]{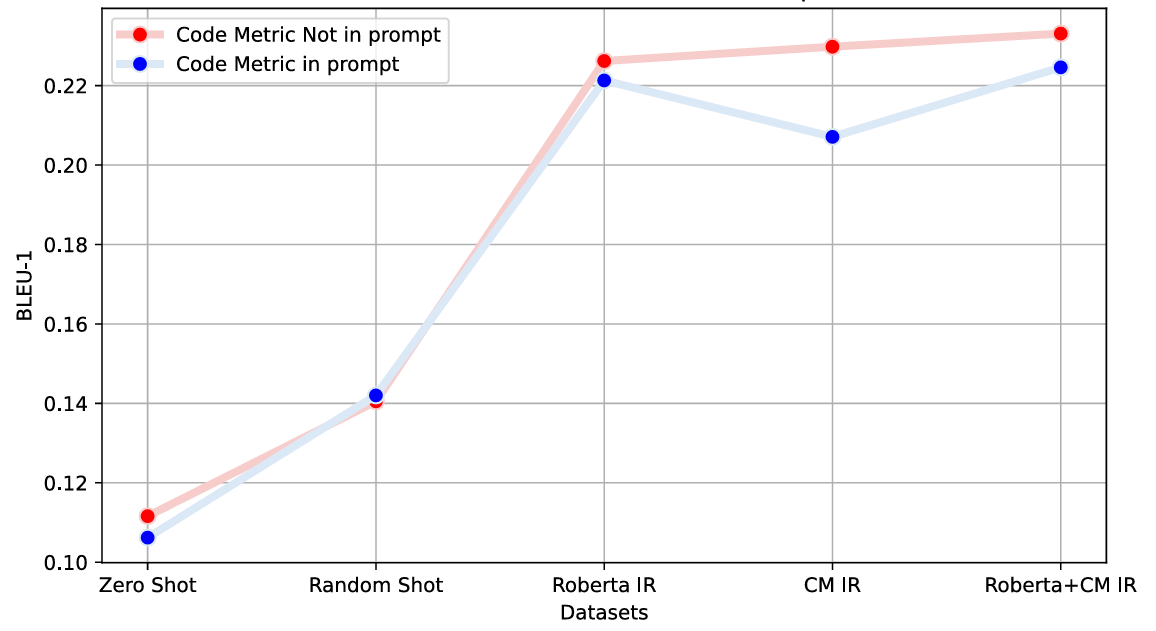}
    \newline
    {\footnotesize (a) BLEU Scores}
  \end{minipage}
  \hfill
  \begin{minipage}[b]{0.32\linewidth}
    \centering
    \includegraphics[width=\linewidth]{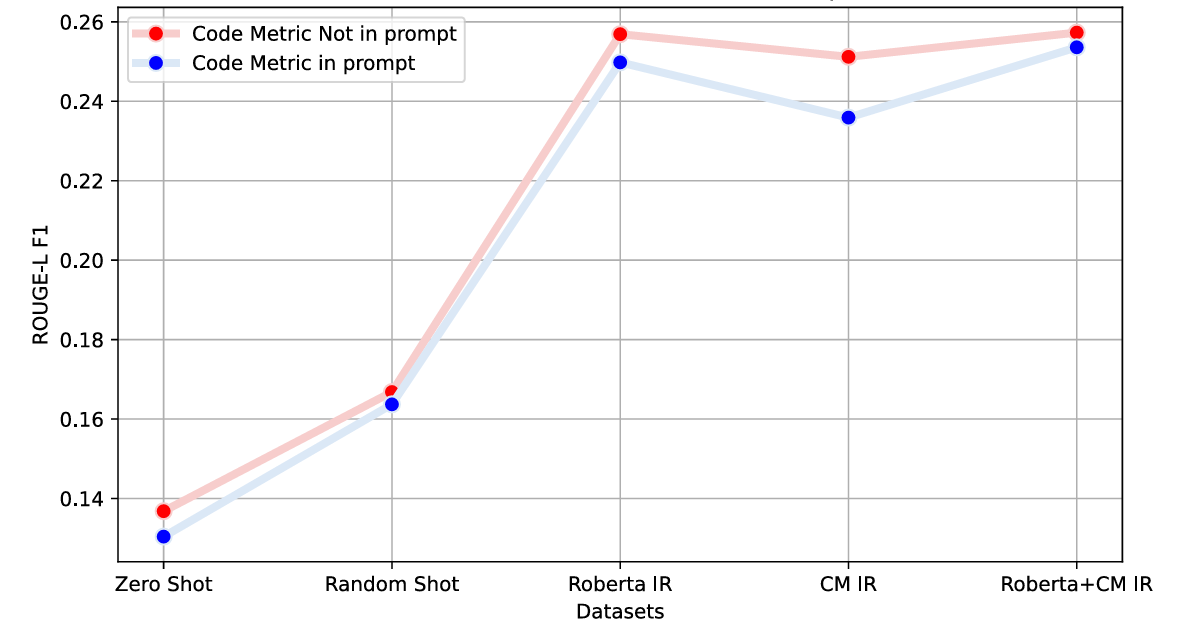}
    \newline
    {\footnotesize (b) ROUGE Scores}
  \end{minipage}
  \hfill
  \begin{minipage}[b]{0.32\linewidth}
    \centering
    \includegraphics[width=\linewidth]{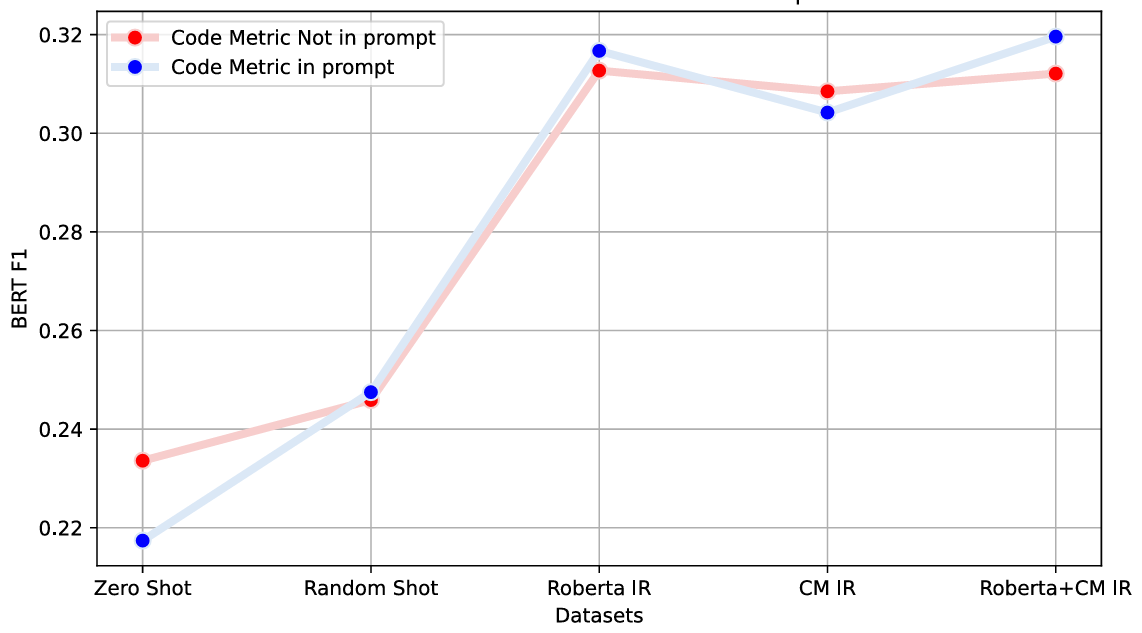}
    \newline
    {\footnotesize (c) BERTScores}
  \end{minipage}
  \caption{Overall Evaluation Results of the GPT‑3.5 Model Across BLEU, ROUGE, and BERTScore Metrics.}
  \label{fig:fewshot_evaluation_result_gpt3_chart}
\end{figure*}

For our LLM experiments conducted with the GPT-3.5 model, we report the BLEU, ROUGE, and BERTScore results for each experiment in \tabref{fewshot_evaluation_result_gpt3_bleu}, \tabref{fewshot_evaluation_result_gpt3_rouge}, and \tabref{fewshot_evaluation_result_gpt3_bert}, respectively. These results correspond to different combinations of the \textit{Shot Sampler} and \textit{Prompt Generator} modules. In addition, \figref{fewshot_evaluation_result_gpt3_chart} illustrates the BLEU-1 score, ROUGE-L F1 measure, and BERTScore F1 measure across all experiments.

As presented in \tabref{fewshot_evaluation_result_gpt3_bleu}, \tabref{fewshot_evaluation_result_gpt3_rouge}, and \tabref{fewshot_evaluation_result_gpt3_bert}, and summarized in \figref{fewshot_evaluation_result_gpt3_chart}, the incorporation of code metrics leads to performance improvements over the baseline Zero-Shot and Random-Shot settings in both configurations. Nonetheless, the results indicate that utilizing code metrics to identify and select the most relevant shots for inclusion in the prompt yields superior outcomes compared to directly embedding the metrics within the prompt. This finding suggests that the shot sampling process constitutes a more effective mechanism for leveraging code metrics.

\begin{tcolorbox}
\textbf{RQ2:} \textit{How can code metrics be effectively integrated into LLM prompts to enhance the accuracy, relevance, and fluency of generated code documentation?}
\par \noindent
\textbf{Answer:} We combined five shot sampling techniques (Zero-Shot, Random-Shot, CM~IR, RoBERTa~IR, and CM+RoBERTa~IR) with two types of prompt generation, with and without code metrics. Our evaluation using GPT-3.5 shows that incorporating code metrics within the shot sampling process, particularly in the \textit{CM~IR} module, is more effective than embedding them directly in the prompt. This finding suggests that code cells with similar quality metrics tend to produce comparable documentation, thereby enhancing the model's learning efficiency.
\end{tcolorbox}

\begin{tcolorbox}
\textbf{RQ3:} \textit{Can the code metrics improve the performance of LLMs in documentation generation for computational notebooks?}
\par \noindent
\textbf{Answer:} Our experiments with GPT-3.5 demonstrate that the proposed \textit{Shot Samplers}, namely \textit{CM~IR} and \textit{RoBERTa+CM~IR}, consistently outperform the \textit{Zero-Shot} and \textit{Random-Shot} baselines. Furthermore, given their comparable performance to the \textit{RoBERTa~IR} approach, we identify the \textit{CM~IR Shot Sampler} as a strong candidate for few-shot learning applications, particularly because it attains similar accuracy while operating entirely on CPU.
\end{tcolorbox}

\subsection{LLM-as-a-Judge Evaluation}
\label{sec:llm-evaluation}
Instead of relying on human evaluation, this study adopts an \textit{LLM-as-a-Judge} paradigm inspired by~\cite{sun2025source} and further supported by recent findings in~\cite{crupi2025effectiveness}, wherein a modern LLM automatically assesses documentation quality. The work in~\cite{sun2025source} demonstrates that GPT-4's evaluation scores are highly correlated with expert ratings ($\rho > 0.9$), confirming its reliability as a scalable alternative to manual evaluation. Similar to human evaluation, for each sample, the evaluator LLM receives three inputs: (i) the code snippet to be summarized, (ii) the reference (original) summary, and (iii) candidate summaries generated by different approaches. GPT-4 is then instructed to rate each candidate summary across three core quality dimensions:

\begin{itemize}[leftmargin=14pt]
    \item \emph{Content Accuracy:} Evaluates how precisely the generated summary reflects the functionality and intent of the original code, ensuring semantic fidelity.
    \item \emph{Fluency and Conciseness:} Assesses the linguistic naturalness and brevity of the summary, penalizing redundancy and awkward phrasing.
    \item \emph{Comprehension Support:} Measures the extent to which the summary aids a developer in quickly understanding the code's logic and purpose.
\end{itemize}

Each aspect is independently rated by the LLM on a 1--5 scale, and the overall score for each summary is computed as the mean of its three ratings. To ensure determinism and reproducibility, the temperature parameter of the LLM evaluator was fixed at $T = 0$. In total, the LLM judge evaluated 1,000 samples. As noted by~\cite{crupi2025effectiveness}, the accuracy and consistency of LLM-based judgments tend to decline for functions that are more complex, require deeper reasoning over code logic, or involve ambiguous natural language descriptions. To address this limitation, a complementary human validation stage was introduced. A random subset of 100 samples was independently reviewed by three expert software engineers, each with over five years of professional experience, to ensure robustness against potential model misjudgments in high-complexity cases. The LLM-based and human-based evaluations exhibited a strong positive correlation ($\rho = 0.88$, $\kappa = 0.84$), confirming the reliability of the GPT-4-based judging framework for this task.

\begin{table}[t]
\centering
\caption{LLM-as-a-Judge Evaluation Results for CM~IR, Roberta~IR, and Zero-Shot Approaches}
\label{tab:llm-evaluation-results}
\begin{tabular}{|>{\centering\arraybackslash}m{1.6cm}|
                    >{\centering\arraybackslash}m{1.7cm}|
                    >{\centering\arraybackslash}m{2.1cm}|
                    >{\centering\arraybackslash}m{2.5cm}|
                    >{\centering\arraybackslash}m{1.4cm}|}
\hline
\textbf{Approach} & 
\makecell{\textbf{Content}\\\textbf{Accuracy}} & 
\makecell{\textbf{Fluency \&}\\\textbf{Conciseness}} & 
\makecell{\textbf{Comprehension}\\\textbf{Support}} & 
\makecell{\textbf{Overall}\\\textbf{Score}} \\ \hline
Zero-Shot & 3.65 & 3.55 & 3.48 & 3.56 \\ \hline
Random Shot  & 4.04 & 3.96 & 4.08 & 4.02 \\ \hline
Roberta~IR & 4.25 & 4.18 & 4.20 & 4.21 \\ \hline
CM~IR & 4.03 & 4.44 & 4.20 & 4.22 \\ \hline
CM~IR + Roberta IR & 4.33 & 4.29 & 4.27 & 4.30 \\ \hline
\end{tabular}%
\end{table}

As illustrated in \tabref{llm-evaluation-results}, the \textit{LLM-as-a-Judge} framework yields consistent and discriminative results across all evaluated approaches. Both metric-enhanced methods, namely \textit{CM~IR} and \textit{RoBERTa~IR}, clearly outperform the \textit{Zero-Shot} and \textit{Random-Shot} baselines in every evaluation dimension, demonstrating the positive impact of incorporating code metrics into the summarization process. Moreover, the combined configuration, \textit{CM~IR+RoBERTa~IR}, achieves the highest overall performance, highlighting the complementary strengths of these two approaches. Manual inspection of representative samples further revealed that selecting shots based on code-metric similarity produces behavior more consistent with human authors. For instance, well-known libraries tend to receive concise or minimal descriptions, whereas lesser-known libraries and more complex code blocks are documented in greater detail.

\section{Threats to Validity}\label{sec:threads}

This section outlines the potential threats to validity and limitations of the findings, as well as the approaches taken to mitigate them.

From the perspective of \textit{internal validity}, the implementation process may introduce potential threats. Although high-quality and widely used libraries such as PyTorch and HuggingFace were employed, several modules, including the CNN–RNN network structure, were independently implemented. As a result, the evaluation results may have been influenced by development errors. To mitigate this risk, two authors of the paper independently reviewed and verified the source code to ensure correctness and consistency. Moreover, since the results align with prior research, particularly in terms of BLEU and ROUGE scores, and the zero-shot approach performing worse than the few-shot approach, it is believed that possible implementation bugs have not substantially affected the validity of the findings.

In terms of \textit{external validity}, the dataset was exclusively built from Kaggle notebooks, which are generally well-documented and relevant to data science. However, results obtained from such data may not generalize to notebooks from other platforms or to tasks beyond the data science domain. The dataset’s focus on a specific sequential type of code cells and related markdowns also introduces limitations, as portions of the markdown data might not fully correspond to their associated code cells, making the findings less generalizable to other code structures or documentation styles. Furthermore, GPT-3.5 was chosen for evaluation to ensure that the model had not been exposed to the dataset beforehand; however, this choice may itself introduce a threat to validity, as the use of a specific model could limit the generalizability of the results to other language models.

Regarding \textit{construct validity}, the study assumes that traditional deep-learning models do not inherently capture the structural information represented by code metrics. Although this assumption lacks prior empirical evidence and may not hold in all contexts, the evaluation demonstrates that incorporating metrics improves documentation generation accuracy, indicating that the chosen constructs partially reflect the intended phenomena. Furthermore, the effect of each metric individually, as well as other possible metric types, was not examined, which can limit the strength of the construct-level interpretation.

Finally, in terms of \textit{replicability}, both the dataset and source code employed for training and evaluation have been made publicly available~\cite{ReplicationPackage}. This transparency allows other researchers to reproduce the experiments and verify the reported results, thereby reducing potential threats to replicability and enhancing the overall trustworthiness of the findings.

\section{Related Work}
\label{sec:related_work}
This section reviews prior work on automated code documentation.

\subsection{Code Documentation Approaches}
Early efforts in automated code documentation, like~\cite{sridhara2010towards, badihi2017crowdsummarizer}, primarily relied on static code analysis, which often failed to capture contextual dependencies between program elements. The work in~\cite{mcburney2014automatic} extended this line of research by integrating contextual analysis within a natural language generation (NLG) framework~\cite{reiter1997building}, leveraging SWUM representations~\cite{hill2009automatically} and PageRank-based importance measures~\cite{langville2006google} to improve the relevance of generated summaries.
The introduction of neural machine translation (NMT) architectures~\cite{bahdanau2014neural} marked a major advancement in dynamic and semantic code understanding. Sequence-to-sequence models with LSTM-based encoders~\cite{hu2018deep, liu2018table}, graph-based encoders such as Graph2Seq~\cite{xu2018graph2seq}, tree-based models like Code2Seq~\cite{alon2018code2seq}, and other deep neural network-based approaches, like~\cite{AGHAMOHAMMADI2020}, significantly enhanced summarization quality, though at substantial computational cost.

With the emergence of LLMs, code summarization evolved through zero-shot and few-shot prompting~\cite{khan2022automatic, nashid2023retrieval}. Studies have shown that enriching prompts with semantic and structural metadata, such as repository names, function signatures, and data-flow graphs (DFGs), enhances contextual relevance and coherence~\cite{ahmed2024automatic, ahmed2022learning, ahmed2022few, mastropaolo2024evaluating}. Retrieval-augmented prompting further strengthens generation quality by grounding outputs in external knowledge. Specialized Code-LMs such as CodeBERT~\cite{feng2020codebert}
and CodeT5~\cite{wang2021codet5} were designed to overcome the limitations of general-purpose LLMs, while subsequent adaptations, such as CodeBERTER~\cite{saberi2023model}, multilingual lightweight fine-tuning~\cite{saberi2023multilingual, saberi2024utilization}, and hybrid systems like ADAMO~\cite{gu2022assemble}, demonstrate the advantages of combining syntactic and semantic representations. Intent-aware models~\cite{mu2023developer, geng2024large} extended summarization to capture developer rationale, and self-reflective prompting~\cite{ahmed2024automatic} improved adaptability across code complexities.

Recent research has emphasized the integration of structural and semantic cues to enhance documentation quality beyond surface-level syntax. Techniques leveraging ASTs~\cite{alon2018code2seq}, API-level transfer learning~\cite{hu2018summarizing}, and data-flow graph embeddings, as in GraphCodeBERT~\cite{guo2021graphcodebert}, have improved both code comprehension and summarization accuracy. Similarly, prompt formulations infused with semantic metadata, such as repository or function-level context~\cite{ahmed2024automatic}, yield more coherent and meaningful outputs.

Overall, while these advancements collectively demonstrate the benefits of structurally and semantically enriched representations, most prior work remains limited to standalone code files and does not address the fragmented, interleaved, and context-rich nature of computational notebooks. This gap motivates the development of documentation models specifically tailored to notebook environments, where code, text, and execution flow are tightly coupled.

\subsection{Documenting Computational Notebooks}
Computational notebook documentation has recently attracted growing attention due to its critical role in ensuring reproducible research. Rule-based approaches such as~\cite{venkatesh2021automated} rely on static analysis but exhibit limited generalization across diverse notebook structures. Deep learning systems like Themisto~\cite{wang2022documentation} integrate automatic and user-driven components, achieving superior performance compared to traditional models such as Code2Seq. To address the limitations of static and monolithic designs, subsequent studies introduced cell-level structural modeling. HAConvGNN~\cite{liu2021haconvgnn} incorporated hierarchical attention over preceding cells and AST features, while Cell2Doc~\cite{mondal2023cell2doc} employed modular documentation at the pipeline level. These approaches highlight the challenge of capturing inter-cell dependencies and contextual metadata, motivating the integration of LLM-based semantic and usability cues~\cite{sun2025source, mastropaolo2024evaluating}.

Overall, recent research underscores the inherent complexity of notebook documentation and advances toward modular, structure-aware, and multi-modal learning strategies. In our work, we further incorporate code metrics to better understand and represent the structural properties of code.

\section{Conclusions and Future Work}
\label{sec:conclusion}
This study investigated the role of code metrics in enhancing automated documentation generation for computational notebooks using both traditional deep learning models and modern LLMs. Experimental results demonstrated that incorporating code metrics consistently improved BLEU and ROUGE scores, indicating better alignment between generated and reference documentation.
For LLM-based experiments, integrating code metrics within the IR process, particularly through the proposed \textit{CM~IR} and \textit{Roberta+CM~IR} modules, produced superior results compared with Zero-Shot and Random-Shot baselines. The \textit{CM~IR~Shot~Sampler} further showed strong potential as an efficient few-shot approach that can operate effectively on CPU resources.

Overall, the findings confirm that code metrics serve as a lightweight yet impactful enhancement for documentation generation systems. Future research will explore the development of more advanced and context-sensitive code metrics, deeper integration of structural and semantic features, and the application of these methods to broader software engineering tasks to further advance automation and documentation quality.

\bibliographystyle{ACM-Reference-Format}
\bibliography{references.bib}

\end{document}